\documentclass{elsart}
\usepackage{amssymb}
\usepackage{amsmath}
\usepackage{graphics}

\usepackage{epsfig}
\usepackage{amssymb}

\def\bfg #1{{\mbox{\boldmath $#1$}}}

\begin{document}

\begin{frontmatter}

\title{
A possibility to determine the P-parity of the $\Theta^+$
pentaquark in the  ${ p}{ n}\to  \Lambda^0\Theta^+$ reaction
}

\author[dubna]{Yu.N.~Uzikov}
\footnote{E-mail address: uzikov@nusun.jinr.ru
}

\address[dubna]{Joint Institute for Nuclear Research, LNP, 141980 Dubna,
 Moscow Region, Russia}


\begin{abstract}
{ Spin structure of the reaction ${\vec p}{\vec n}\to {\vec \Lambda^0}
{\vec \Theta}$ is analyzed at the threshold in a model independent way 
 under assumption that the $\Theta^+$ is an isosinglet.
 We found that the sign of the spin-spin 
 correlation parameter $C_{x,x}$ being measured in a double-spin experiment,
 determines the P-parity of the $\Theta^+$ unambiguously. Furthermore we 
 show that the polarization coefficients $K_x^x,K_y^y$ and 
 $K_z^z$ which describe the 
 polarization transfer  
 from  polarized beam or target to the final $\Lambda^0$ and $\Theta^+$
 are nonzero for a positive parity of the $\Theta^+$
 and equal zero for a negative parity. It allows one to determine
 the P-parity of the  $\Theta^+$  in a single-spin measurement,
 since the polarization of the $\Lambda^0$ can be measured  via its decay
 $\Lambda^0\to \pi^-+ p$.}
\end{abstract}  

\begin{keyword}
Pentaquark, strangeness, spin observables

\begin{PACS}
13.75.Cs, 14.20-c\\[1ex]
\end{PACS}
\end{keyword}

\end{frontmatter}
\baselineskip 4ex
\newpage

 The  recent experimental discovery of an exotic barion with
 a positive strangeness $S=+1$ 
 and surprisingly  narrow width 
\cite{Nakano,Barmin,Stepanyan,Kubarovsky,Elsa,SVD}, 
 called now as the  $\Theta^+(1540)$,
 stimulated many theoretical works concerning its structure.
 The quantum numbers of this barion such as spin, parity and  isospin
 are not yet  determined experimentally. According to the original
 prediction
 within   the chiral soliton model \cite{diakonov},
  the pentaquark $\Theta^+$ belongs
  to the anti-decuplet with all members  having one the same spin-parity, 
  namely $J^P=\frac{1}{2}^+$.  
  From the point of view of constituent quark model the minimal 
  number  of quarks in the $\Theta^+$ is five, i.e. the quark content
 of this
  barion is  $uudd{\bar s}$. Within the naive picture with  noninteracting
  quarks, the ground state of the $\Theta^+$ is
  expected to be the S-state, therefore the P-parity of the $\Theta^+$ 
  has to
  be negative, $P=-1$.  
  Inclusion of the special type of $qq$-interaction into the quark model
  could lead to the positive parity \cite{hosaka,glozman}. 
  Diquark model \cite{jaffe} predicts also $P=+1$. On the other hand, 
  the Lattice QCD calculation  \cite{lattice}
  predict  for this barion
  $P=-1$. Therefore, for quark dynamics the P-parity of the $\Theta^+$
  is a key point and it has to be determined experimentally.
  
  Several methods depending on the dynamical assumptions where
  suggested for determination of the
 P-parity of the $\Theta^+$ \cite{theory}.  
  According to general theorem \cite{moravchik}, in order to determine the
  parity of one particle in a binary reaction $1+2 \to 3+4$ one has to know
 polarizations at least of two fermions participating in this reaction.
 Model independent methods for determination of the P-parity of the
 $\Theta^+$  were suggested recently in Ref. \cite{thomas} and more detail 
 in  Ref. \cite{hanhart} for pp-collision and in Ref.\cite{rekalo}
 for photoproduction  of the $\Theta^+$. 
The method of Refs.\cite{thomas,hanhart} 
 is based on measurement of the spin-spin
 correlation parameter in the reaction ${\vec p}{\vec p} \to 
\Sigma^+\theta^+$  near the threshold.
 We show  here that the reactions
 ${\vec p}{\vec n}\to \Lambda^0\Theta^+ $ and    
 ${\vec p}{ n}\to {\vec \Lambda^0} \Theta^+ $ can be also  used for
 the  P-parity 
 determination in a  model-independent way under assumption that the isospin
 of the $\Theta^+$ is known.
 We assume  conservation of the P-parity,  total
 angular momentum and isospin in the reaction and use  the
  generalized Pauli principle for nucleons.
  We assume here that $\Theta^+$ is an isosinglet, since the isospin
  partner  $\Theta^{++}$ was  not observed in $\gamma p$  interaction 
 \cite{Elsa,Kubarovsky}.
  At last, we assume that the spin of the $\Theta^+$ is
  $\frac{1}{2}$.  Some remarks related to the isospin T=1 and
   spin $J=\frac{3}{2}$ for
  the $\Theta^+$ are given at the end of the paper.

  Under assumption that the $\Theta^+$ is an isosinglet,
  the total isospin of the initial $pn$ state equals
  zero in the rection $pn\to \Lambda^0\Theta^+$. Furthermore,
  according to generalized
  Pauli principle, the orbital momentum $L$  of the $pn$ system is 
  even for
  the spin triplet state S=1 and odd for the singlet state, S=0.
  We consider  here the threshold region with an
  excess energy less than few tens MeV. At this
  condition the S-wave dominates in the final state \cite{hanhart,Nam2004}.
    Using P-parity and total angular momentum conservation, one can find
  that   for $P=-1$  of the $\Theta^+$ there is only one 
  transition, i.e. $^1P_1\to~^3S_1$
  and for $P=+1$ there is only the  $^3S_1-~^3D_1\to~^3S_1$ transition.
  We discuss these transitions  below separately.

 {\bf Negative parity.}
  In nonrelativistic formalism,
  the matrix element for the $^1P_1\to ^3S_1$ transition
 can be written   as
  \begin{equation}
   \label{matrelemp}
    F=({\bf T}^\prime \cdot {\bf k}) S\, f,
\end{equation}
  where  $f$ is a complex amplitude,
 ${\bf T}^\prime=i(\chi_{\sigma_3}^+\,{\bfg \sigma}
{\sigma_y}\,
\chi_{\sigma_4}^{(T)+})$,
$S=-i(\chi_{\sigma_2}^T\,
\sigma_y \,\chi_{\sigma_1})$,
${\bfg \sigma}$ is the Pauli spin matrix, $\chi_{\sigma_j}$ is the
 Pauli spinor
for the j-th particle with the spin projection $\sigma_j$,
 and ${\bf k}$ is the unit vector along the beam direction.
  The polarized  cross section for this transition is
  \begin{equation}
  \label{polarizedp}
  d\sigma^{neg}({\bf p}_1,{\bf p}_2)=
 d\sigma_0^{neg}(1-{\bf p}_1\cdot {\bf p}_2),
 \end{equation} 
 where ${\bf p}_j$ is the polarization of the beam or target $(j=1,2)$,
 $d\sigma_0^{neg}$ is the unpolarized cross section.
 Below  we use for  spin-observables the notations  
 defined in \cite{ohlsen} and assume  that the $OZ$ axis is directed 
 along the vector {\bf k}.
 As can be seen  from Eq. (\ref{polarizedp}), the spin-spin correlation
  parameters in the initial state equal to $-1$: 
 $ C_{xx}=C_{yy}=C_{zz}=-1$.  We can show  using Eq. (\ref{matrelemp})
 that   the spin transfer coefficients
 are zeros,  $K_j^i=0, \,(i,j =x,y,z)$.

{\bf Positive parity}. 
 The only matrix element
 for this case,  $^3S_1-~^3D_1\to~^3S_1$, can be written as 
 \begin{equation}
  \label{matrelemsd}
 F = G\, ({\bf T}^\prime \cdot {\bf T})+ 
({\bf T}^\prime \cdot {\bf k})({\bf T}\cdot {\bf k})\,F,
 \end{equation}
 where 
${\bf T}=-i(\chi_{\sigma_2}^T \sigma_y\,{\bfg \sigma}\,
\chi_{\sigma_1})$ and G and F are complex amplitudes.
 This amplitudes can be
written also as 
$G= U-W/\sqrt{2},\, F=3/\sqrt{2}\,W $, where $U$ and $W$ are the  
 S- and 
D-waves, respectively, (see, for example, \cite{alkhazov})
 in the initial state.
 For polarized beam and target, the  cross section takes the following
 form  
\begin{equation}
\label{crossp1p2plus}
d\sigma^{pos}({\bf p}_1,{\bf p}_2)= d\sigma_0^{pos}\left 
\{1+A({\bf p}_1\cdot {\bf p}_2) +B ({\bf k}\cdot {\bf p}_1)
({\bf k}\cdot {\bf p}_2)\right \}.
\end{equation}
 Here $d\sigma_0^{pos}$ is the spin-averaged cross section,
 which can be written as 
\begin{equation}
\label{averagedsd}
 d\sigma_0^{pos}= \frac{1}{4} K\,\left \{ |G+F|^2 + 2|G|^2 \right \},
\end{equation}
where $K$ is the kinematical factor. The factors $A$ and $B$ in
 Eq.(\ref{crossp1p2plus}) have 
 a  form
\begin{eqnarray}
\label {a}
A=\frac{|G+F|^2}{|G+F|^2+2|G|^2}, \\
B=-2\frac{|F|^2+2ReGF^*}{ |G+F|^2+2|G|^2}.
\label{b}
\end{eqnarray}
One can see from
Eq.(\ref{crossp1p2plus}) that non-zero spin-spin correlation parameters 
 are the following
\begin{eqnarray}
\label{spi-spin-sd}
C_{x,x}=C_{y,y}=A, \\
\label{czz}
C_{z,z}= A+B=\frac{|G-F|^2 - 2|F|^2}{|G+F|^2+2|G|^2}.
\end{eqnarray}
 As it follows from Eqs. (\ref{a},\ref{spi-spin-sd}),
 the coefficients $C_{x,x}$ and  $C_{y,y}$ are positive for $P=+1$.
 On the other hand, these observables are  negative
 and maximal in absolute value 
 for $P=-1$ (see Eq.(\ref{polarizedp})).
    This result does not depend on the mechanism
  of the reaction   and therefore allows one to determine the P-parity
  unambiguously in double-spin measurements with transversely
 polarized  beam
  and target. This result is similar to that found recently for
  the spin-spin correlation in the reaction
  ${\vec p}{\vec p}\to  \Sigma^+\theta^+$ near threshold 
\cite{thomas,hanhart}.
  The sign of the coefficient $C_{z,z}$ given by Eq.(\ref{czz}) 
  can   be positive or negative for $P=+1$, depending on 
  the relative weight
  of the S- and D- waves in this transition. 
   
Furthermore, in case of $P=+1$ we found  the spin  transfer
 coefficients  as
\begin{eqnarray}
\label{Kij}
K_x^x=K_y^y =2\frac{|G|^2+Re GF^*}{|G+F|^2+2|G|^2},\\ \nonumber
K_z^z=  2\frac{|G|^2}{|G+F|^2+2|G|^2},\\ \nonumber
K_x^y=K_x^z=K_y^x=K_y^z=K_z^x=K_z^y=0.
\end{eqnarray}
 At last, we can show that for unpolarized beam (or target), the polarization
 of the final particles is zero  in the reaction 
$pn\to \Lambda^0\Theta^+$ independently on the sign of the P-parity
of the $\Theta^+$ and  the analyzing power is  zero also.

 As follows from Eqs. (\ref{Kij}), 
 for polarized beam (or target) the
 final particle is polarized along the direction of the initial
 polarization
 vector,
 if the P-parity of the $\Theta^+$ is positive.  
 The sign and the absolute value of the spin-transfer coefficients
 depends on the relative strength
 of the S- and D-component and therefore can not be calculated
 without further  dynamical assumptions.
 For the negative
 parity $P=-1$ of the $\Theta^+$ the polarization transfer from the 
 beam (or target) to the final particle is zero.
 Therefore, a measurement of the polarization of one final particle 
 in the reactions ${\vec p}n\to \Lambda^0 +\Theta^+$ or
 $ p{\vec n}\to \Lambda^0 +\Theta^+$  is equivalent to determination of
 the P-parity of the $\Theta^+$ in a largely model-independent way
\footnote{
 We assume here that  there is no
 cancellation between the S- and D-waves ($ U\not =W/\sqrt{2}$)
  at the threshold of this reaction
  and therefore $G\not =0$.}.
 The polarization
 of the final $\Theta^+$ is hardly be measured, but a measurement
 of the polarization of the  $\Lambda^0$ is possible by measurement of
 the
 angular distribution  in the decay $\Lambda^0\to \pi^-+ p$.
 Indeed, due to P-parity violation in this decay, there is a 
 large asymmetry in angular distribution of final particles in the 
  c.m.s. 
 of the $\Lambda^0$ in respect of the
  direction of the $\Lambda^0$ spin.   
 At some experimental conditions  a such single-spin experiment
 is, probably,  more
  simple than the double-spin measurement in the ${\vec p}{\vec p}
 \to \Sigma^+\Theta^+$ or ${\vec p}{\vec n}\to \Lambda^0\Theta^+$
 reactions. At present, a such measurement is possible at COSY
 in reaction ${\vec p}d \to \Lambda^0+\Theta^++ p_{sp}$ with
 polarized proton  beam and unpolarized
 deuteron target in a region of quasi-free ${\vec p}n$ interaction.
 At low  momenta of the spectator proton $p_{sp}$ less than
 $\approx 50$MeV/c,
 the excess energy in the reaction $pn\to\Lambda^0\Theta^+$ is less than
 50 MeV  that provides the S-wave  dominance in the final state
 \cite{hanhart,Nam2004}. Furthermore, as known from  study of the
 $d(p,2p)n$ reaction \cite{punjabi},  an influence of initial and final
 state
 interactions on spin observables  is rather weak in quasi-free region. 

 Let us make some further remarks. 
 (i) Since the polarization of the $\Sigma^+$ is also self-analyzing
    via its decays   $\Sigma^+\to p+\pi^0$ or $\Sigma^+\to n+\pi^+$,
  we  discuss  here  briefly the polarization transfer 
 in the reaction
  ${\vec p}p\to{\vec \Sigma^+}\Theta^+$. It is easy to show that for
  $P=+1$
  there is no polarization transfer in this reaction, since the 
 spin-singlet transition
 $^1S_0\to~^1S_0$  dominates: $K_i^j=0, \, (i,j=x,y,z)$. 
 For $P=-1$ there are here two transition amplitudes
 \cite{thomas}: $^3P_0 \to ~^1S_0$ and  $^3P_1 \to~^3S_1$
 We found, that 
 the polarization transfer is zero for the
 first transition.  For the spin-triplet transition 
 $^3P_1 \to ^3S_1$ the polarization transfer is given by the
 coefficient $K_z^z=+1$, whereas all others coefficients $K_i^j$ are zero.
 Due to mixing of these two transitions 
 the total $K_z^z$ can be changed  but 
 not vanished, on the whole,   $K_z^z\not =0$.
 Therefore, a measurement of the longitudinal 
  polarization of the $\Sigma^+$ in the
 reaction ${\vec p}p\to{\vec \Sigma^+}\Theta^+$ with longitudinally
  polarized
 beam can be used as a filter for the
 P-parity of the $\Theta^+$.
   
 (ii) We can show also that for polarized beam (or target) there
 are spin-spin correlations in the final state of the reaction
 $pn\to \Lambda^0\Theta^+$
 for $P=+1$ and no correlations 
 for $P=-1$. However, we do not discuss 
 these effects in this note, because to observe them experimentally
 one has to measure polarizations of the $\Lambda^0$ and $\Theta^+$
 simultaneously that seems unlikely at present.  

(iii) {\bf If the isospin of the  $\Theta^+$ is  equal to 1},
 then the total isospin of the initial  $pn$ system is $I=1$ 
in the reaction $pn\to
\Lambda^0\Theta^+$. In this case one has the same transition
 amplitudes and, therefore, {\it the same spin observables
  as in the reaction $pp\to\Sigma^+\Theta^+$}. As follows from
Refs. \cite{thomas,hanhart} and above discussion, the spin observables
 in the reaction $pp\to\Sigma^+\Theta^+$ are essentially different
 as compared to the reaction
 $pn\to \Lambda^0\Theta^+$ (with the isosinglet $\Theta^+$), namely:
%
  $C_{x,x}=-1$ and $K_i^j=0$ $(i,j=x,y,z$) for $P=+1$,
 whereas  the $C_{x,x}$ is nonnegative \cite{hanhart} and 
 $ K_z^z \not  =0$, if  $P=-1$.
 On the contrary,
 the spin observables of the reaction ${\vec p}{\vec p}
  \to\Sigma^+\Theta^+$
 are not sensitive to  the isospin of the $\Theta^+$,
 when it takes the possible values $0,1$ or $2$.
  Obviously, the reaction $pn\to \Lambda^0\Theta^+$ is forbidden for
 the isospin $T=2$ of the $\Theta^+$ due to isospin invariance of
 strong interactions.
 
(iv) 
 In chiral  models (see, for example, \cite{glozman}),
 the $\Theta^+ (\frac{1}{2}^P)$ could have a partner with 
 the spin $J=\frac{3}{2}$.
 For the spin  $\frac{3}{2}$ of the $\Theta^+$, at the threshold 
 of the reaction $pp\to\Sigma^+\Theta^+$ there is
 one transition $^1D_2\to~^5S_2$ for $P=+1$ and
 two transitions $^3P_1\to~^3S_1$ and  $^3P_2-~^3F_2\to~^5S_2$
 for  $P=-1$. Since  for $P=+1$ there is only one amplitude
 with the spin-siglet initial state, one has got for this case
 $C_{x,x}=C_{y,y}=C_{z,z}=-1$. For $P=-1$, the initial state is
 the spin-triplet, therefore one should expect   that $C_{x,x}$ is
  nonnegative, $0\leq C_{x,x} \leq +1$.
 In the reaction $pn\to \Lambda^0\Theta^+$ with the spin
 $J=\frac{3}{2}$  of the $\Theta^+$, we have  the same
 transitions  as for $J=\frac{1}{2}$,  but only in the case 
 of $P=+1$ one new transition,
  $^3D_2\to ~^5D_2$,  contributes 
  in addition to the $^3S_1-~^3D_1\to~^3S_1$ 
 transition. Since the all  transitions for $P=+1$  are  the 
 spin-triplet ones in the $pn$ initial state,  one should  expect
 that the spin-spin correlation parameter  $C_{x,x}$ 
 is still nonnegative for $P=+1$, whereas for $P=-1$ we found 
 $C_{x,x}=C_{y,y}=C_{z,z}=-1$. Therefore, in both these reactions, $pp-$
 and $pn-$,  the sign of the spin-spin
 correlation parameter $C_{x,x}$ allows one to determine the P-parity of the
 $\Theta^+$ unambiguously for both cases  $J=\frac{1}{2}$ and $J=\frac{3}{2}$.
 A question about  the polarization transfer
 coefficients for $J=\frac{3}{2}$ is more complicated and 
 will be studied separately.

 In conclusion, assuming that the $\Theta^+$ is the isosinglet with the
 spin $\frac{1}{2}$,  we have analyzed  in a model independent way
 the spin-spin correlation parameters $C_{i,j}$ and spin-transfer 
 coefficients $K_i^j$ of the reaction $pn\to \Lambda^0\Theta^+$ near
 the threshold. We found  
 that the P-parity of the $\Theta^+$ can be measured in a single spin 
 experiment with the transversally or longitudinally  polarized beam or 
 target if 
 the polarization of the final $\Lambda^0$ is  measured via the decay 
 $\Lambda^0\to \pi^-+p$.  A similar result is found here for the
 ${\vec p}p\to {\vec \Sigma^+}\Theta^+$ with the longitudinally 
 polarized beam. In contrast to the reaction
 $ p\, p\to\Sigma^+\Theta^+$,
  in  the reaction $p\, n\to \Lambda^0\Theta^+$ the sign of  $C_{x,x}$
 coincides with the sign of the P-parity of the $\Theta^+$ and the
 non-zero polarization transfer occurs only for $P=+1$. 
 However, if the  $\Theta^+$ is   an  isotriplet, 
 the spin observables of the  reaction $pn\to \Lambda^0\Theta^+$ 
 are identical with those
 for the reaction $pp\to \Sigma^+\Theta^+$.
 Therefore a measurement
 of the above spin observables in these two reactions allows one
 to determine both the P-parity and isospin of the $\Theta^+$.

\end{document}